\documentclass[twocolumn,aps,showpacs]{revtex4}
\usepackage[latin9]{inputenc}
\usepackage{amsmath}
\usepackage{amssymb}
\usepackage{graphicx}

\makeatletter

\providecommand{\tabularnewline}{\\}

\@ifundefined{textcolor}{}
{%
 \definecolor{BLACK}{gray}{0}
 \definecolor{WHITE}{gray}{1}
 \definecolor{RED}{rgb}{1,0,0}
 \definecolor{GREEN}{rgb}{0,1,0}
 \definecolor{BLUE}{rgb}{0,0,1}
 \definecolor{CYAN}{cmyk}{1,0,0,0}
 \definecolor{MAGENTA}{cmyk}{0,1,0,0}
 \definecolor{YELLOW}{cmyk}{0,0,1,0}
}

\usepackage{color}

\def\NOT(#1,#2){\OneQubitGate(#1,#2){$X$}}

\makeatother

\begin{document}
\title{Extending and measuring dephasing times of nuclear spins in NV centers
of diamond }
\author{Jingfu Zhang$^{1,2}$, Swathi S. Hegde$^{1,*}$, Fedor Jelezko$^{2,3}$
and Dieter Suter$^{1}$\\
 $^{1}$Fakultaet Physik, Technische Universitaet Dortmund, D-44221
Dortmund, Germany}
\author{$^{2}$Institute for Quantum Optics, Ulm University, 89081 Ulm, Germany}
\author{$^{3}$Center for Integrated Quantum Science and Technology (IQST),
Ulm University, 89081 Ulm, Germany}
\date{\today}
\begin{abstract}
Long coherence times rank among the most important performance measures
for many different types of quantum technology. In NV centers of diamond,
the nuclear spins provide particularly long dephasing times. However,
since initialization and readout require assistance from the electron
spin, the apparent dephasing times can be reduced by the electron
spin lifetime. Here we propose and implement schemes for measuring
and extending the dephasing times of nuclear spins, resulting in dephasing
times that are longer than the longitudinal relaxation time $T_{1}$
of the electron spin.
\end{abstract}
\maketitle
%

\textit{Introduction}.--In all areas of quantum technologies like
quantum computing and quantum sensing \cite{RevModPhys.89.035002},
the preservation of quantum information in the presence of a noisy
environment is one of the most challenging tasks. Hybrid systems such
as nitrogen-vacancy (NV) centers in diamond are promising for implementing
quantum computing \cite{nielsen,Stolze:2008xy,ladd2010quantum,Suter201750,kurizki2015quantum},
which is believed to have a large potential for speeding up the solution
of certain problems that have no easy solutions for classical computers.
One of the main benefits of the NV system lies in the long coherence
time of the nuclear spins, e.g., up to one minute at low temperature
(3.7 K) \cite{PhysRevX.9.031045}.

However, the preservation and measurement of the coherence, are limited
by the longitudinal relaxation time ($T_{1}^{e}$) of the electron
spin \cite{dobrovitski2012,PhysRevA.87.012301,PhysRevA.97.062330}.
Specifically, if the electron spin changes its state as a result of
longitudinal ($T_{1}^{e}$) relaxation, the connected nuclear spin
experiences a different hyperfine coupling, which results in the loss
of its coherence. In addition, the detection of the coherence of the
nuclear spin depends on the polarization of the electron spin and
the loss of this polarization due to longitudinal relaxation reduces
the measured coherence time \cite{dobrovitski2012,PhysRevA.87.012301}.
In this letter, we relax these limitations in two ways: During the
coherence storage period, we partially refocus the nuclear spin coherence
with a Hahn echo. In addition, we modify the detection scheme: instead
of measuring the coherence of the nuclear spin in only one subspace
of the electron state \cite{dobrovitski2012,Suter201750}, we exchange
the states of nuclear spin and the electron spin using a SWAP gate
\cite{nielsen}, which transfers the nuclear spin coherence into electronic
spin coherence that can be measured through the combination of a $\pi/2$
microwave (MW) pulse and a readout laser pulse. In contrast to earlier
work \cite{doi:10.1126/science.1220513} on weakly coupled systems,
or based on encoded qubits \cite{Pfender2017,PhysRevX.12.011048},
our current work focuses on a strongly coupled system without ancillary
qubits.

\textit{System}.--We consider a $^{13}$C nuclear spin strongly coupled
to the electron spin of a single NV center. Here, strong coupling
means that the hyperfine coupling is close to the nuclear spin Larmor
frequency \cite{zhang18,swathi19}. Initially, we do not consider
the $^{14}$N nuclear spin of the NV center but focus onto the subsystem
where the $^{14}$N is (and remains) in the $m_{N}$=1 state. This
is a safe assumption since the population lifetime ($T_{1}^{N}$)
of the $^{14}$N nuclear spin is much longer than that of the electron
spin \cite{Pfender2017,Soshenko_2021}. The spins interact with a
weak magnetic field $B\approx148$ G oriented along the symmetry axis
of the NV center. In a suitable reference frame, we can write the
relevant part of the Hamiltonian as 
\begin{eqnarray}
\frac{1}{2\pi}\mathcal{H}_{e,C} & = & DS_{z}^{2}-(\nu_{e}-A_{N})S_{z}\nonumber \\
 &  & -\nu_{C}I_{z}+A_{zz}S_{z}I_{z}+A_{zx}S_{z}I_{x},
\end{eqnarray}
where $S_{z}$ denotes the electron spin-1 operator, and $I_{x/z}$
the $^{13}$C spin-1/2 operators. The zero-field splitting is $D=2.87$
GHz. $\nu_{e/C}=\gamma_{e/C}B$ denote the Larmor frequencies of the
spins specified by the indices for the electron and $^{13}$C nuclear
spins, with $\gamma$ being the gyromagnetic ratio. $A_{N}=-2.16$
MHz is the secular part of the hyperfine coupling with the $^{14}$N
nuclear spin while $A_{zz}$ and $A_{zx}$ are the relevant components
of the $^{13}$C hyperfine tensor. We measured $\nu_{C}=0.158,$ $A_{zz}=-0.152$
and $A_{zx}=0.110$ MHz \cite{zhang18}. These values are highly suitable
for indirect control of the nuclear spin \cite{PhysRevA.76.032326,PhysRevA.78.010303,swathi19,PhysRevLett.130.090801}.

In the experiment, we used a $^{12}$C enriched diamond as the sample,
which provides a long coherence time of the electron spin ($T_{2}^{*e}\approx35$
$\mu$s). The experiment was performed at room temperature. In the
following, we write $|m_{S}\rangle$ for the eigenstate of $S_{z}$
with eigenvalue $m_{S}$, and $|\uparrow\rangle$ and $|\downarrow\rangle$
for the eigenstates of $I_{z}$ with eigenvalues $\pm1/2$.

To prepare the system for the measurement, we use a combination of
laser and MW pulses \cite{swathi19,zhang18} to initialize the system
into the state

\begin{equation}
\rho_{ini}^{e,C}=s_{1}|0\uparrow\rangle\langle0\uparrow|+(1-s_{1})|1\downarrow\rangle\langle1\downarrow|\label{eq:iniC}
\end{equation}
with $s_{1}=0.80$. The first term is used to generate the coherence
of the nuclear spin. The second term results from the imperfection
in polarizing the $^{13}$C spin and contributes observable effects
in the detected signals. Details of the pulse sequence for preparing
$\rho_{ini}^{e,C}$ are given in the supplementary material (SM, Sect.
IIB). 
\begin{figure}[h]
\centering{}\includegraphics[width=0.66\columnwidth]{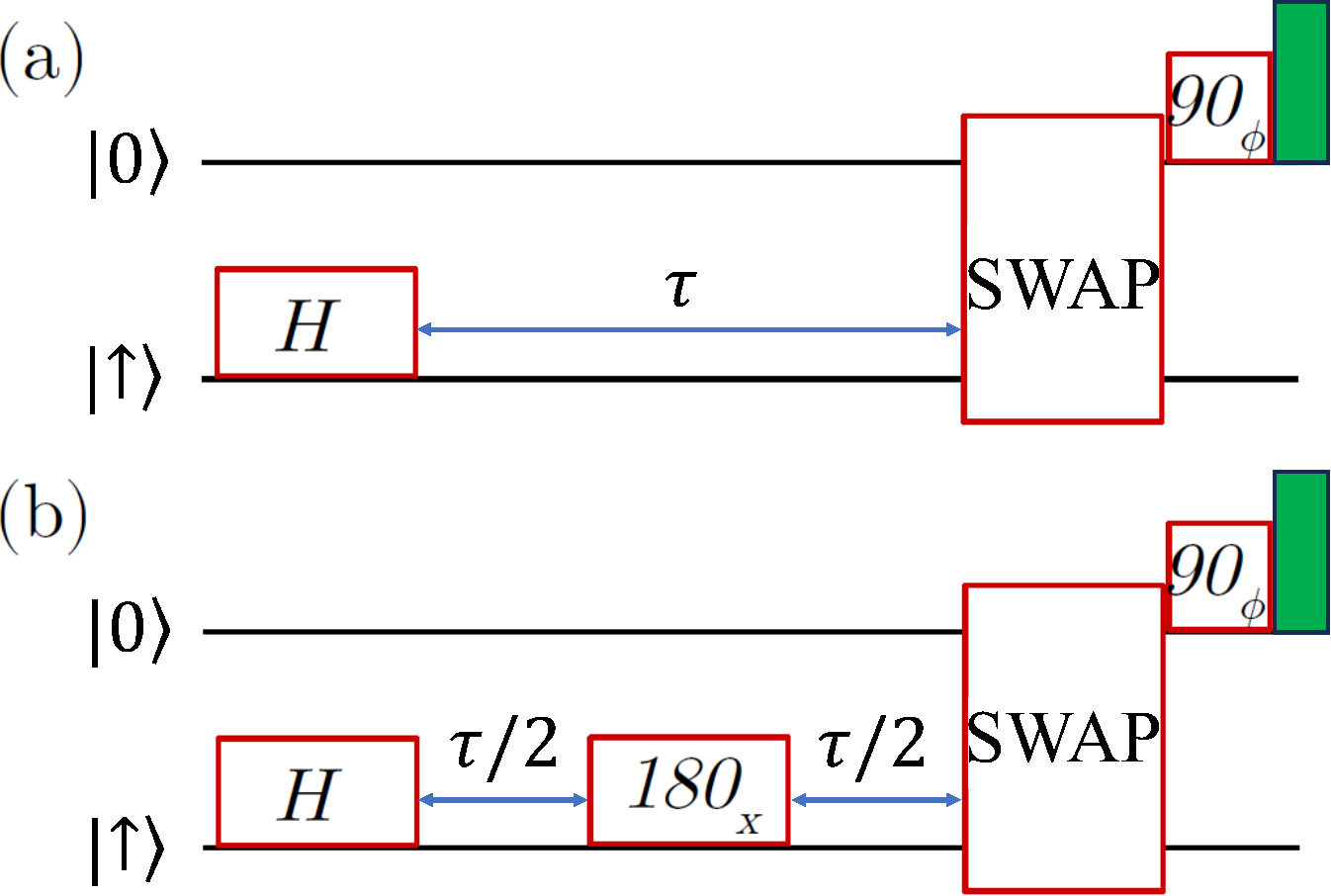}
\caption{Quantum circuits for measuring the dephasing of the $^{13}$C spin
during free precession (a) and with refocusing by a Hahn echo (b).
In each circuit, the top line denotes the electron spin qubit and
the bottom the $^{13}$C spin. The initial state is $|0,\uparrow\rangle$.
$H$ represents a Hadamard gate, $180_{x}$ the refocusing pulse and
SWAP the 2-qubit SWAP gate. The filled rectangles indicate the readout
laser pulses. \label{figcirEC}}
\end{figure}

\begin{figure}[h]
\centering{}\includegraphics[width=1\columnwidth]{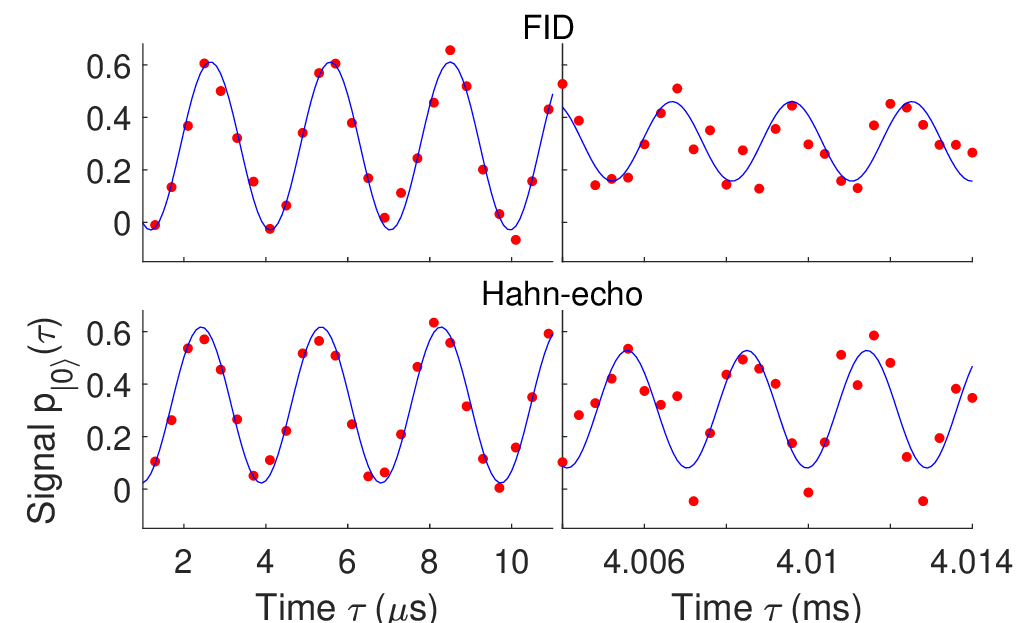}\caption{Measured signals with the fitted curves for the FID and Hahn-echo.
\label{figsignal}}
\end{figure}

\textit{Experimental protocol and results}.--Fig. \ref{figcirEC}
shows the protocol for the relaxation measurement of the $^{13}$C
spin as quantum circuits for the free induction decay (FID) (a) and
the Hahn-echo (b). Here, we only consider the 4-level subsystem with
the initial state corresponding to the first term in Eq. \eqref{eq:iniC}.
In this subsystem, we use unitary operations that correspond to a
Hadamard gate $H$ and refocusing pulse $180_{x}$ to $^{13}$C spin,
and the two-qubit SWAP gate. All these operations were implemented
without RF pulses, using indirect control for the $^{13}$C spin \cite{swathi19,zhang18}.
The final $90_{\phi}$ readout pulse converts a component of the electron
spin coherence into observable population of the electron spin. Its
phase $\phi$ is incremented in proportion to the free precession
time $\tau$: $\phi=2\pi\nu_{d}\tau+\phi_{0}$, where $\phi_{0}$
denotes an initial phase. As a result, the observed signal oscillates
at frequency $\nu_{s}=\nu_{C}+\nu_{d}$ ( $\nu_{d}$) in the FID (Hahn-echo)
experiment. The detuning frequency $\nu_{d}$ was chosen as $\nu_{d}^{a}=-0.5$
MHz in (a) and as $\nu_{d}^{b}=\nu_{d}^{a}+\nu_{C}$ in (b), respectively,
so that the measured signals from circuits (a) and (b) have the same
frequency $\nu_{s}=0.342$ MHz, to allow easier comparison, as illustrated
in Figure \ref{figsignal}. The state of the electron qubit can be
detected by laser pulses, as shown in Fig. \ref{figcirEC} \cite{Suter201750}.

Applying the Hadamard gate $H$ to the $^{13}$C spin generates the
superposition state $|+\rangle=(|\uparrow\rangle+|\downarrow\rangle)/\sqrt{2}$
from $|\uparrow\rangle$. Therefore $\rho_{ini}^{e,C}$ shown in Eq.
(\ref{eq:iniC}) is transformed to 
\begin{equation}
\rho_{sup}^{e,C}=s_{1}|0\rangle\langle0|\otimes|+\rangle\langle+|+(1-s_{1})|1\downarrow\rangle\langle1\downarrow|,\label{eq:stateH-1}
\end{equation}
 where the state of the $^{13}$C spin in the subspace of electron
state $|1\rangle$ remains unchanged by the Hadamard gate $H$. We
discuss the cases of FID and Hahn echo in the following.

In the FID experiment shown in Fig. \ref{figcirEC}(a), the nuclear
spin coherence is allowed to undergo free evolution for a time $\tau$.
The coherence evolves continuously and dephases under the effect of
environmental perturbations. These perturbations include the longitudinal
relaxation of the electron spin ($T_{1}^{e}$): If it flips between
states $|0\rangle$ and $|-1\rangle$ ( $|1\rangle$), the nuclear
spin coherence prepared between $|0,\uparrow\rangle\leftrightarrow|0,\downarrow\rangle$,
with precession frequency $\nu_{0}=\nu_{C}=0.158$ MHz, is transferred
to the $|-1,\uparrow\rangle\leftrightarrow|-1,\downarrow\rangle$
($|1,\uparrow\rangle\leftrightarrow|1,\downarrow\rangle$) transition,
where it precesses with a different frequency $\nu_{-1}=\sqrt{A_{zx}^{2}+(\nu_{C}+A_{zz})^{2}}=0.110$
MHz ($\nu_{1}=\sqrt{A_{zx}^{2}+(\nu_{C}-A_{zz})^{2}}=0.329$ MHz).
Since this transfer is an incoherent process, the coherence of the
nuclear spins coupled to the $m_{S}=\pm1$ electron spin subsystems
vanishes.

We denote the populations of the different electron spin levels as
$P_{m_{S}}(\tau)$, and write the state at the end of the evolution
period as

\begin{eqnarray}
\rho_{FID}^{e,C}(\tau) & = & \sum_{m_{S}}P_{m_{S}}(\tau)(|m_{S}\rangle\langle m_{S}|)\otimes\varrho_{FID}^{C_{m_{S}}}(\tau)\label{eq:state_FID-1}
\end{eqnarray}
where $\varrho_{FID}^{C_{m_{S}}}(\tau)$ denotes the nuclear spin
state in the subspace $|m_{S}\rangle$ of the electron spin, and
\begin{equation}
P_{0}(\tau)=\frac{1}{3}-2(\frac{1}{6}-\frac{s_{1}}{2})e^{-3\kappa\tau}\label{eq:p0T1-1-1}
\end{equation}
\begin{equation}
P_{-1}(\tau)=\frac{1}{3}+(\frac{1}{6}-\frac{s_{1}}{2})e^{-3\kappa\tau}-\frac{1}{2}(1-s_{1})e^{-\kappa\tau}\label{eq:pm1T1-1-1}
\end{equation}
\begin{equation}
P_{1}(\tau)=\frac{1}{3}+(\frac{1}{6}-\frac{s_{1}}{2})e^{-3\kappa\tau}+\frac{1}{2}(1-s_{1})e^{-\kappa\tau}
\end{equation}
with $T_{1}^{e}=1/(3\kappa)$, where $T_{1}^{e}$ is the electron
spin relaxation time and its experimental value is 5.5 ms. Details
are in Section I of the SM. The state of the nuclear spin in the electron
spin subspace $|1\rangle$ does not contribute to the observed signals,
since the readout is selective for the subspace ${|0\rangle,|-1\rangle}$.
In the subspace $m_{S}=0$ we have with 
\begin{eqnarray}
\varrho_{FID}^{C_{0}}(\tau) & = & \frac{1}{2}\left(\begin{array}{cc}
1 & c_{FID}(\tau)e^{-i2\pi\nu_{C}\tau}\\
c_{FID}(\tau)e^{i2\pi\nu_{C}\tau} & 1
\end{array}\right).\label{eq:FID}
\end{eqnarray}
In the following, we assume an exponential decay of the coherence
$c_{FID}(\tau)=e^{-\tau/T_{2}^{*C}}$. More details on Eq. \eqref{eq:FID}
are presented in the SM. The populations $P_{\pm1}(\tau)$ of the
electron spin state $|\pm1\rangle$ are the result of incoherent ($T_{1}^{e}$)
process, and therefore no nuclear spin coherence builds up in these
subsystems.

To generate an observable signal from the nuclear spin coherence,
we use a 3-step process: we first apply a SWAP operation to the 2-qubit
system to convert the nuclear spin coherence into electron spin coherence,
then we apply a $90^{\circ}$ pulse to the electron and count the
fluorescence photons during the readout-laser-pulse \cite{Suter201750}.

The SWAP operation exchanges the states of the $^{13}$C nuclear spin
qubit with that of the electron spin qubit. It transforms the state
$\rho_{FID}^{e,C}(\tau)$ in Eq. (\ref{eq:state_FID-1}) to 
\begin{eqnarray}
\rho_{FID,S}^{e,C}(\tau) & = & P_{0}(\tau)\varrho_{FID}^{C_{0}}(\tau)\otimes(|\uparrow\rangle\langle\uparrow|)\nonumber \\
 & + & P_{-1}(\tau)E/2\otimes(|\downarrow\rangle\langle\downarrow|)\nonumber \\
 & + & P_{1}(\tau)|1\rangle\langle1|\otimes\varrho_{FID}^{C_{1}}(\tau).\label{eq:roh_swap-2}
\end{eqnarray}
Here, $\varrho_{FID}^{C_{0}}$ is defined in Eq. \eqref{eq:FID},
and $E$ denotes the $2\times2$ unit operator. The SWAP operation
has therefore transferred the coherence $c_{FID}(\tau)$ to the electron
spin states. From there, it is converted into a population difference
by a 90$^{\circ}$ MW pulse with a time-dependent phase $\phi=2\pi\nu_{d}^{a}\tau+\phi_{0}$.
Finally, a readout laser pulse measures the population of the electron
spin in the state $|0\rangle$ , 
\begin{eqnarray}
p_{|0\rangle,FID}(\tau) & = & (1/2)[P_{0}(\tau)+P_{-1}(\tau)\nonumber \\
 & + & P_{0}(\tau)c_{FID}(\tau)\sin(2\pi\nu_{s}\tau+\phi_{0})],\label{eq:sim_C.m-2}
\end{eqnarray}
where $\nu_{s}=-(\nu_{C}+\nu_{d}^{a})=0.342$ MHz. Here and in the
following, $p$ refers to the population after the SWAP operation
and the 90$^{\circ}$ pulse, and $P$ to the population before the
SWAP. Figure \ref{figsignal} illustrates the measured $p_{|0\rangle,FID}(\tau)$
for two time segments. More results are listed in the SM. 

\begin{figure}[h]
\centering{}\includegraphics[width=1\columnwidth]{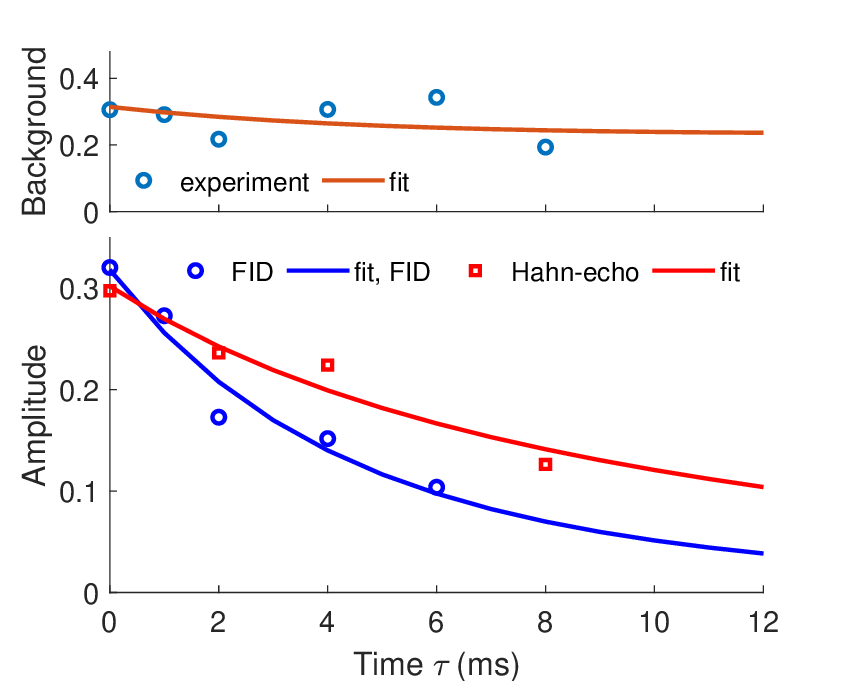}
\caption{Experimental results for the background $[P_{0}(\tau)+P_{-1}(\tau)]/2$
(top) and for the oscillation amplitude (bottom) in the FID and Hahn-echo
experiments, which measures the nuclear spin coherence. \label{figResEC}}
\end{figure}

The signal $p_{|0\rangle,FID}(\tau)$ in Eq. (\ref{eq:sim_C.m-2})
consists of an oscillating part that reflects the coherence in the
system, with amplitude $P_{0}(\tau)c_{FID}(\tau)/2$ and a background
$[P_{0}(\tau)+P_{-1}(\tau)]/2$, which reflects the populations. Fig.
\ref{figResEC} shows the experimental results, where we fit the measured
background in $p_{|0\rangle,FID}(\tau)$ to the function $[P_{0}(\tau)+P_{-1}(\tau)]/2-d_{0}$
with the offset $d_{0}=0.086$. More details are presented in the
SM. By fitting the measured amplitude of the coherence $p_{|0\rangle,FID}(\tau)$
to the function $P_{0}(\tau)c_{FID}(\tau)/2$, where
\begin{equation}
c_{FID}(\tau)=c_{0,FID}e^{-\tau/T_{2}^{*C}},\label{fit-1-2}
\end{equation}
we obtain the dephasing time $T_{2}^{*C}$. The experimental results
and the corresponding fits are shown in Fig. \ref{figResEC} (bottom),
where the amplitude denotes $P_{0}(\tau)c_{FID}(\tau)/2$ or $[P_{0}(\tau)+P_{-1}(\tau)]c_{Hahn}(\tau)/2$
for FID or Hahn-echo experiment, respectively. Additional details
are given in the SM. In Table, \ref{params_fit-1}, we list the obtained
parameters.

\begin{table}
\begin{tabular}{|l|c|c|}
\hline 
$^{13}$C, FID & $c_{0,FID}=0.80$ & $T_{2}^{*C}=8.66$ ms\tabularnewline
\hline 
$^{13}$C, Hahn-echo & $c_{0,Hahn}=0.76$ & $T_{2}^{C}=14.10$ ms\tabularnewline
\hline 
\end{tabular}

\caption{Parameters obtained by fitting the experimental data in Figs. \ref{figResEC}
to the function \eqref{fit-1-2}. The details on the normalization
of the experimental data are presented in Section II of the SM.}
\label{params_fit-1} 
\end{table}
The Hahn-echo experiment starts with the same initialization as the
FID experiment but uses an inversion pulse in the middle of the evolution
period to partly refocus the evolution, as shown in Fig. \ref{figcirEC}(b).
Additional details are given in the SM. The state at the qubit system
at the end of the full evolution ($\frac{\tau}{2}-180_{x}-\frac{\tau}{2}$)
and the SWAP operation is 
\begin{eqnarray*}
\rho_{Hahn,S}^{e,C}(\tau) & = & \varrho_{Hahn}(\tau)\otimes\left(\begin{array}{cc}
P_{0}(\tau) & 0\\
0 & P_{-1}(\tau)
\end{array}\right)\\
 & + & P_{1}(\tau)|1\rangle\langle1|\otimes\varrho_{Hahn}^{C_{1}}(\tau),
\end{eqnarray*}
 where the operator 

\begin{eqnarray*}
\varrho_{Hahn}(\tau) & = & \frac{1}{2}\left(\begin{array}{cc}
1 & c_{Hahn}(\tau)\\
\\c_{Hahn}(\tau) & 1
\end{array}\right)\\
c_{Hahn}(\tau) & = & e^{-\tau/T_{2}^{C}}.
\end{eqnarray*}
is now an electron spin operator in the (0,-1) subspace. 

Similar to the detection in the FID experiment, we apply a 90$^{\circ}$
pulse with a phase $\phi=2\pi\nu_{d}^{b}\tau+\phi_{0}$ to obtain
the population
\begin{eqnarray}
p_{|0\rangle,Hahn}(\tau) & = & [P_{0}(\tau)+P_{-1}(\tau)]/2\nonumber \\
 & \cdot & [1+c_{Hahn}(\tau)\sin(2\pi\nu_{d}^{b}\tau+\phi_{0})].\label{eq:sim_C.m-2-1}
\end{eqnarray}
Here we choose the detuning frequency as $\nu_{d}^{b}=-0.342$ MHz,
different from the case of the FID, so that the resulting oscillation
frequencies are the same. The experimental data can then be fitted
to $[P_{0}(\tau)+P_{-1}(\tau)]c_{Hahn}(\tau)/2$, where 
\begin{equation}
c_{Hahn}(\tau)=c_{0,Hahn}e^{-\tau/T_{2}^{C}}.\label{fit-1-2-1}
\end{equation}
This allows us to extract $T_{2}$. Table \ref{params_fit-1} lists
the resulting parameters. They show that the refocusing pulse in the
Hahn echo extends the dephasing by 60\% compared to the FID.

Eq. (\ref{eq:sim_C.m-2-1}) shows, that of $T_{1}^{e}$ still limits
the amplitude of the detected signals in measuring the coherence of
the nuclear spin, but the effect is smaller than in the case of the
FID experiment, see Eq. (\ref{eq:sim_C.m-2}). Applying echo pulses
to the nuclear spin in the subspace of $m_{S}=1$ might further relax
the limitation.

\textit{Conclusion and outlook}.-- We have measured the dephasing
of the $^{13}$C nuclear spin in a single NV center during free precession
and with partial refocusing in a Hahn echo experiment. Compared with
previous works, e.g. \cite{dobrovitski2012,PhysRevA.87.012301}, we
introudced an improved detection scheme for the nuclear spin coherence,
which relies on a SWAP gate. In the case of the FID, the nuclear spin
dephasing time $T_{2}^{*}$ exceeds $T_{1}^{e}$ by a factor of 1.6.
Using one refocusing pulse, we further extended the coherence time
to 2.6 times $T_{1}^{e}$. The effect of electron spin flips between
states $|0\rangle$ and $|-1\rangle$ due to $T_{1}^{e}$ relaxation
can be suppressed in the following way: In the subspace $|0\rangle$,
the echo can refocus the evolution of the nuclear spin, while in subspace
$|-1\rangle$, the nuclear spin can be approximated by an eigenstate
of the effective Hamiltonian ($\sim I_{x}$) and therefore does not
evolve. In the present work, we only use one echo on the nuclear spin
in a subspace of two states $|0\rangle$ and $|-1\rangle$ of the
electron spin, where indirect control in this subspace. We believe
that multiple echos should extend the coherence time further \cite{RevModPhys.88.041001,suter2012},
and might provide a useful method to investigate the physics model
for $T_{1}^{e}$, where the nuclear spin works as a useful ancillary
quantum memory \cite{Pfender2017,PhysRevLett.116.230502,Zaiser2016}.
Our experiment was performed in NV centers. The method should also
work in other systems, such as malonic acid radical \cite{PhysRevA.78.010303,PhysRevLett.107.170503}
and other solid defects \cite{PhysRevLett.132.026901}.

\textit{Acknowledgement}.--This project has received funding from
the European Union's Horizon 2020 research and innovation programme
under grant agreement No 828946. The publication reflects the opinion
of the authors; the agency and the commission may not be held responsible
for the information contained in it. J. Z. thanks Nick Grimm for helpful
discussions. JZ acknowledges support by Union\textquoteright s HORIZON
Europe program via project SPINUS (No. 101135699).

{*} Present affiliation: Qruise GmbH, Am Hoelzersbach 7, D-66113 Saarbruecken, Germany.


\end{document}